\documentclass[sigconf]{acmart} 
\AtBeginDocument{%
  \providecommand\BibTeX{{%
    \normalfont B\kern-0.5em{\scshape i\kern-0.25em b}\kern-0.8em\TeX}}}

\setcopyright{acmcopyright}
\copyrightyear{2018}
\acmYear{2018}
\acmDOI{10.1145/1122445.1122456}

\acmConference[AeSIR '21]{AeSIR '21: Automated Support to Improve code Readability}{November 15--11, 2021}{}

\setcopyright{none} 
\renewcommand\footnotetextcopyrightpermission[1]{} 

\begin{document}



\title{Readability and Understandability of Snippets Recommended by General-purpose Web Search Engines: a Comparative Study}




\author{Carlos Eduardo C. Dantas}
\email{carloseduardodantas@iftm.edu.br}
\affiliation{%
  \institution{Federal University of Uberlândia}
  \country{Brazil}
}

\author{Marcelo A. Maia}
\email{marcelo.maia@ufu.br}
\affiliation{%
  \institution{Federal University of Uberlândia}
  \country{Brazil}
}  

\renewcommand{\shortauthors}{Carlos Eduardo C. Dantas and Marcelo A. Maia.}

\begin{abstract}
Developers often search for reusable code snippets on general-purpose web search engines like Google, Yahoo! or Microsoft Bing. But some of these code snippets may have poor quality in terms of readability or understandability. In this paper, we propose an empirical analysis to analyze the readability and understandability score from snippets extracted from the web using three independent variables: ranking, general-purpose web search engine and recommended site. We collected the top-5 recommended sites and their respective code snippet recommendations using Google, Yahoo!, and Bing for 9,480 queries, and evaluate their readability and understandability scores. We found that some recommended sites have significantly better readability and understandability scores than others. The better-ranked code snippet is not necessarily more readable or understandable than a lower-ranked code snippet for all general-purpose web search engines. Moreover, considering the readability score, Google has better-ranked code snippets compared to Yahoo! or Microsoft Bing. 
\end{abstract}



\maketitle

\section{Introduction}

Code snippets (or code examples) are some lines of reusable source code to show how to solve a specific programming problem \cite{Keivanloo2014}. Developers often search for reusable code snippets on the web \cite{Xia2017}, especially on programming sites as StackOverflow \cite{Hucka2016}, or using general-purpose web search engines like Google, Yahoo! or Microsoft Bing to find examples for their respective programming tasks \cite{Rahman2018}. Those code snippets may be straightforwardly reused in software under development, to decrease the time to perform the programming tasks and accelerate the development process \cite{Holmes2013}. Although search engines like Google have over 200 different factors to rank the results \cite{Evans2007AnalysingGR}, the top-ranked pages could have poor quality code examples in terms of readability and reusability features \cite{HORA2021110971}. 

Although Google is the most popular general-purpose web search engine having more than 90\% of market share \footnote{https://gs.statcounter.com/search-engine-market-share}, the other general-purpose web search engines as Yahoo! and Microsoft Bing are available to search code snippets, and the developers could be interested to know how different web search engines ranks top-level readable and understandable code snippets. Another issue consists in evaluating how readability inter-relates to understandability. While readability is associated to reading and comprehending the syntax, understandability is associated to the semantic aspect of code snippets, e.g., the \textit{statements}, \textit{beacons} or \textit{motifs} \cite{Scalabrino2017}. In this context, we propose an investigation to assess how general-purpose web search engines as Google, Microsoft Bing, and Yahoo! rank code examples using features, such as readability and understandability. The study is driven by the following research questions:
 
 \begin{itemize}
     \item \textbf{RQ \#1)} How are code snippets recommended by general-purpose web search engines ranked in terms of readability and understandability? This research question analyses if code snippets with the higher score in readability and understandability features are generally ranked in a specific position in the rank interval [1,5].
     \item \textbf{RQ \#2)} How do general-purpose web search engines compare to each other in terms of readability and understandability features?  This research question compares recommended code snippets by Google, Microsoft Bing, and Yahoo!, to verify if any of these general-purpose web search engines recommend code snippets with the higher score in readability and understandability.
     \item \textbf{RQ \#3)} How do recommended sites containing code snippets compare to each other in terms of readability and understandability features? This research question compares the code snippets from the five most popular sites recommended by the general-purpose web search engines to verify if any site has code snippets with the higher score in readability and understandability.
 \end{itemize}

The paper is organized as follows. Section 2 shows a motivating example. Section 3 discusses the related researches. Section 4 presents the study design proposed to collect the web pages, code snippets, and metrics. The results are reported and discussed in Section 5. Section 6 presents the qualitative discussion about the results. Section 7 has the threats that could affect the validity of this study. And finally, Section 8 summarizes our observations in lessons learned and outlines directions for future work.

\section{Motivating Example}

The motivational example was extracted from Hora's study \cite{Hora2021Google}. In a search for \textit{File.mkdirs} examples in \textit{Google}, the Figures 1 and 2 in \cite{Hora2021Google} 
shows two recommended sites (\textit{Tutorialspoint} and \textit{JavaTutorialHQ})  and their respective code \textit{snippets}. The suggestion from \textit{Tutorialspoint} has worst readability and reusability metric values, but it is better ranked by Google compared to the \textit{JavaTutorialHQ} solution. A possible explanation is because \textit{Tutorialspoint} has natural language explanations similar to the input query. To verify how \textit{Google} rank the code snippets without natural language, \citet{Hora2021Google} has build a \textit{site} with web pages having only code snippets for some tasks. After Google has indexed the new site, the same Google search was performed on that new web site, and then the \textit{JavaTutorialHQ} code snippet was better ranked compared to the \textit{Tutorialspoint} snippet.

The motivational example raises two hypotheses:

 \begin{itemize}
     \item \textbf{H1)} A better-ranked code snippet could not necessarily have higher readability or understandability score.
     \item     \textbf{H2)} Some recommended sites could have overall better readability or understandability score than other sites.
 \end{itemize}    
 


\section{Related Work}

Some works have been proposed to analyze the readability of source code snippets. In a related paper with ours,   \citet{Hora2021Google} investigated how Google is ranking the code snippets in terms of the readability and reusability features. This research constructed a new site with 1,000 web pages with code snippets from 100 Java \textit{APIs}. After Google indexed its web pages, some input queries were performed using Google narrowing that site. The objective was to identify how Google ranked the web pages containing only code snippets. Google does not necessarily prioritize code snippets with high readability or reusability metric values, but web pages with multiple code snippets would probably be ranked first. Our research has differences because we consider other general-purpose web search engines, such as Microsoft Bing and Yahoo!, and also consider the understandability feature.

Other works uses readability feature to improve the code snippet overall score. \citet{Hora2021APISonarMA} constructed the API \textit{Sonar tool}  \footnote{http://apisonar.com/} ranking code snippets with readability feature proposed by \citet{Scalabrino2018}. \citet{Moreno2015} developed the \textit{Muse} approach  to rank code examples using readability feature proposed by  \citet{Buse2010}. These related works shows how readability is a well used feature to rank code snippets. Our work differs because it is not ranking code snippets, but compares code snippets readability and understandability features in three independent variables: ranking, recommended websites, and general-purpose web search engine.

Some related works reinforce the relevance of our study: \citet{Treude2017} shows that only 49\% of StackOverflow code snippets are fully self-explanatory, which explains why developers are interested in explanations accompanying StackOverflow code snippets \cite{Nasehi2012}. Our work helps to compare the readability and understandability scores between StackOverflow accepted answers and other sites, to identify potential sites with more comprehensive code snippets.

\section{Study Design}

This section presents the overall approach to answer the research questions. The major steps are: (1) Select Input Queries, (2) Collect Top-n Web Pages, (3) Extract Code Snippets, (4) Calculate  Metrics, and (5) Analysis Methods. The details of each step are in the following subsections. A replication package, including the tools, scripts, evaluations and the  instructions for  reproduction is available \cite{carlos_eduardo_c_dantas_2021_5224346}.   

\subsection{Select Input Queries}

In this step, we selected 10,000 input queries performed by users on CROKAGE tool \footnote{http://isel.ufu.br:9000/}. CROKAGE is a code search engine that extract code snippets written in Java language and their explanations from \textit{StackOverflow} \cite{DBLP:journals/ese/SilvaRRSPDM20}. These input queries were performed by users from more than 80 countries, searching for programming tasks \footnote{https://stackoverflow.blog/2019/08/14/crokage-a-new-way-to-search-stack-overflow/}. We removed duplicated queries and queries manually labeled as \textit{not applicable} (e.g., non-Java programming languages) by the CROKAGE research \cite{DBLP:journals/ese/SilvaRRSPDM20}.

\subsection{Collect Top-n Web Pages}

\begin{figure*}[!h]
\centerline{\includegraphics[width=0.8\textwidth]{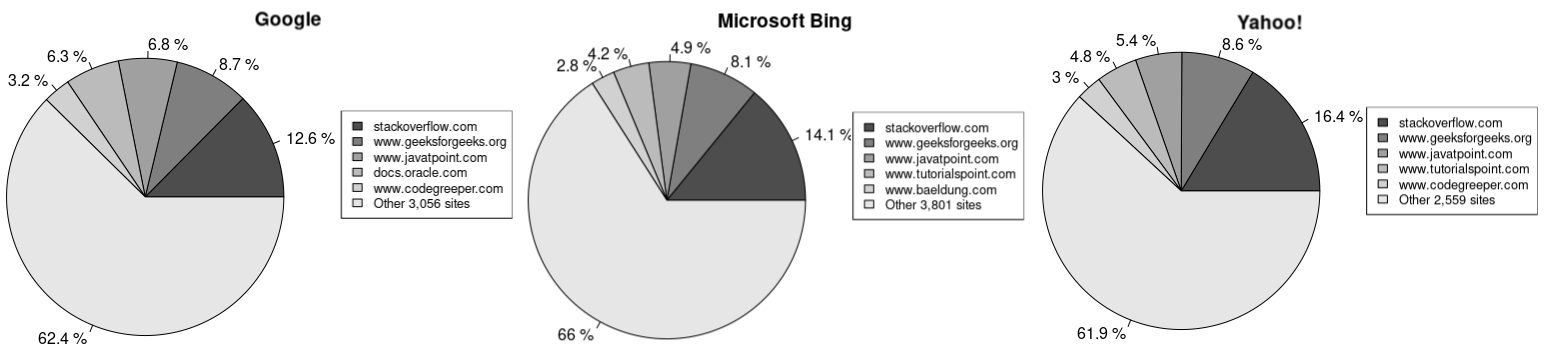}}
\vspace{-4mm}
\caption{Most popular sites with recommended web pages ranked on top-5 for 9,480 queries performed on Google, Microsoft Bing and Yahoo!}
\label{fig:extract}
\end{figure*}

This step consists in collecting the top-5 web pages recommended by Google, Yahoo!, and Microsoft Bing for each input query. These queries receive the additional tokens \textit{"example in java"}. The \textit{"example"} token aims to find code snippets instead of only explanations \cite{Treude2018}. The token \textit{in java} aims to find code snippets written in Java language. General queries as \textit{how to add element in a list} could return code snippets written in other popular programming languages like Python or Javascript \cite{Hora2021Google}. To collect the recommended web pages, we used regular expressions in each general-purpose web search engine, extracting the links to recommend web pages from their \textit{html tags}. We discarded queries with less than 5 web pages recommendations, resulting in 9,480 queries and a total of 47,400 links to web pages from 5,355 distinct sites. Figure \ref{fig:extract} shows the five most popular recommended sites from each web search.

\subsection{Extract Code Snippets}

This step consists in selecting the five most popular sites for the input queries and creating regular expressions to extract the code snippets from their \textit{html tags}. The five popular web sites are:

\begin{itemize}
    \item \textit{stackoverflow.com} 
    \item \textit{www.geeksforgeeks.org} 
    \item \textit{www.javatpoint.com}  
    \item \textit{www.tutorialspoint.com} 
    \item \textit{ www.codegrepper.com} 
\end{itemize}

For \textit{stackoverflow.com}, we extracted the code snippets from each accepted answer written in Java, because the questioner should have the best judgement of whether the answer solves the problem \cite{Yang2016}. In the other sites, the regular expressions search for source code and natural language description with the tokens \textit{"example"} and \textit{"Java"} inside the \textit{html tags}. The \textit{www.javatpoint.com} site has specific \textit{CSS class} for each programming language, making it  easier to find Java source code.

\subsection{Calculate Metrics}

This step consists in calculating the readability and understandability scores on the extracted code snippets. 

To measure readability, this research uses the prediction model proposed by \citet{Scalabrino2018}. This model has been used in other works to extract readability scores from code snippets \cite{Hora2021Google} \cite{Hora2021APISonarMA}, and the classes inside GitHub repositories \cite{Piantadosi2020}. This model\footnote{https://dibt.unimol.it/report/readability/} includes a set of metrics including comments, identifiers consistency, textual coherence, number of meanings, and concepts. The output score is a real number in the interval [0,1] where 0 means \textbf{low readability} and 1 means \textbf{high readability}. 

To measure understandability, the selected code-based metric  is cognitive complexity proposed by  \citet{Campbell2018} with the source code is available in \textit{SonarSource} tool \footnote{https://www.sonarsource.com/docs/CognitiveComplexity.pdf}. The cognitive complexity could be used to measure some understandability aspects \cite{Marvin2020} \cite{Wyrich2021}. The output score is a natural number, where 0 means \textbf{high understandability}, and if the score is equal or higher than 15, is considered \textbf{low understandability} \footnote{https://stackoverflow.com/questions/45083653/sonarqube-qualify-cognitive-complexity/45084107\#45084107}.

To measure understandability in the same interval [0,1] as readability, we propose the metric as follows:

\vspace{-4mm}
\[ understandability(cs_i) =
  \begin{cases}
    1 - \frac{\#cc}{\#mcc}    & \quad \text{if } \#cc < \text{15}\\
    0.0 & \quad \text{otherwise}
  \end{cases}
\]
\vspace{-3mm}

\textit{\#cc} is the complexity cognitive value extracted from \textit{SonarSource} tool for the code snippet \textit{$cs_i$}. The \textit{\#mcc} is the maximum recommend complexity cognitive value, \textit{\#mcc} = 15. If a code snippet reaches \textit{\#cc} >= 15, the metric output will be 0. 

\subsection{Analysis Methods}

To compare the distribution between the dependent variables (readability score, understandability score) and independent variables (general-purpose web search engine, ranking and recommended sites), we apply the analysis of variance (\textit{ANOVA}) using 5\% confident level (i.e., \textit{p-value<0.05}). To find which groups are significantly different from each other, we use \textit{Tukey test}. This analysis have three groups on general-purpose web search engines (Google, Microsoft Bing! and Yahoo), five groups on ranking (top-1 to top-5), and five groups on the selected sites (\textit{stackoverflow}, \textit{geeksforgeeks}, \textit{javatpoint}, \textit{tutorialspoint} and \textit{codegrepper}).

 
 
\section{Results and Findings}

\begin{figure*}[!h]
\centerline{\includegraphics[width=0.8\textwidth]{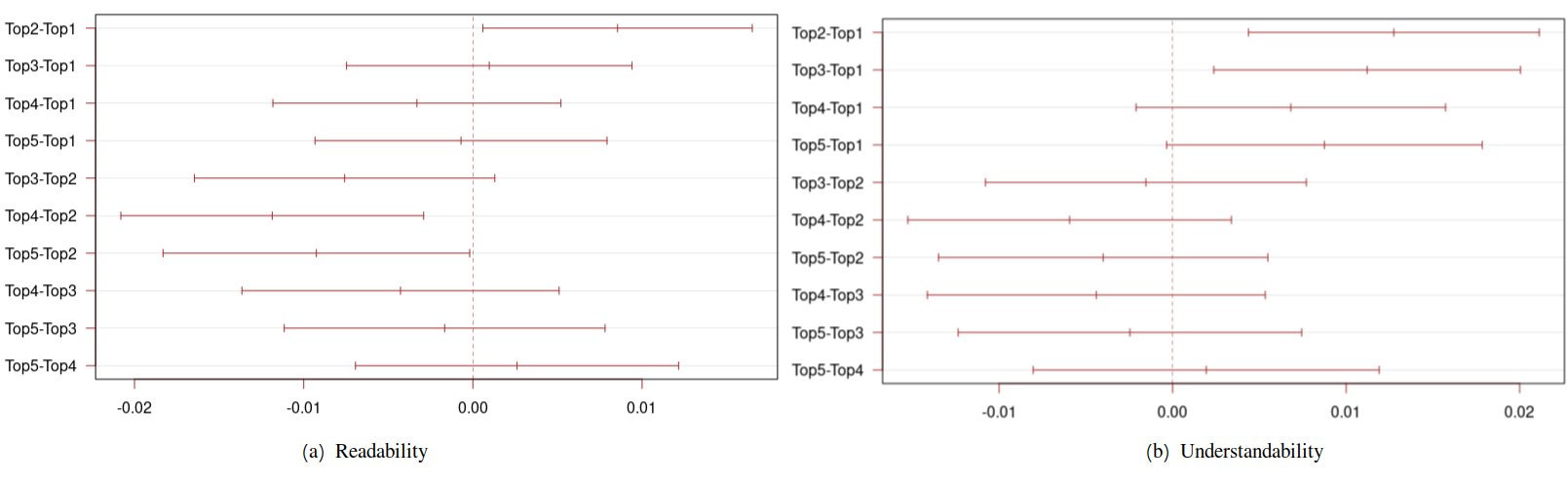}}
\vspace{-4mm}
\caption{\textit{Tukey Test} confidence intervals (x-axis) with the differences between ranking groups (y-axis)}
\label{fig:rq1}
\end{figure*}

\begin{figure*}[!h]
\centerline{\includegraphics[width=0.8\textwidth]{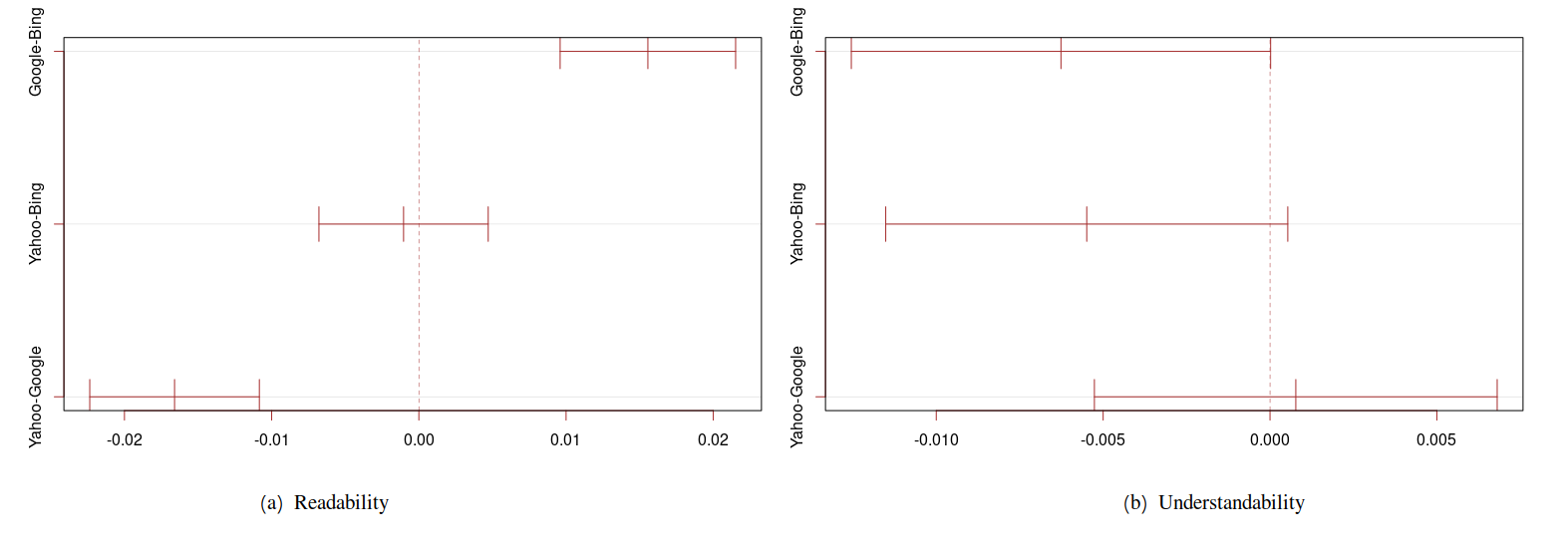}}
\vspace{-4mm}
\caption{\textit{Tukey Test} confidence intervals (x-axis) with the differences between web search engines groups (y-axis)}
\label{fig:rq2}
\end{figure*}

\begin{figure*}[!h]
\centerline{\includegraphics[width=0.8\textwidth]{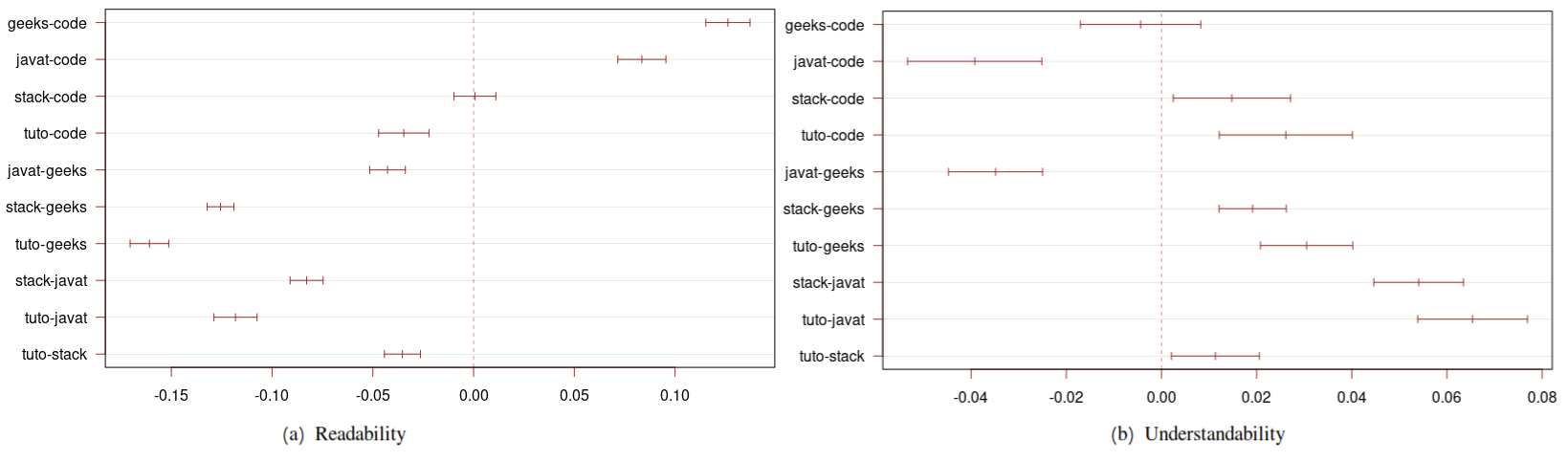}}
\vspace{-4mm}
\caption{\textit{Tukey Test} confidence intervals (x-axis) with the differences between recommended sites groups (y-axis)}
\label{fig:rq3}
\end{figure*}

In this section, the results are shown according to each research question. To interpret the \textit{Tukey test} in Figures \ref{fig:rq1},\ref{fig:rq2} and \ref{fig:rq3}, there is significant difference in mean loss for each group which the 0.00 value is outside of the confidence interval. 

\textbf{RQ \#1) How code snippets recommended by general-purpose web search engines are ranked in terms of readability and understandability?}

Extracting the \textit{ANOVA} on ranking independent variable, we obtained \textit{p-value = 0.0034} for readability and \textit{p-value = 0.0003} for understandability. This result means that there is significant differences between ranking code snippets in readability and understandability. The Figure \ref{fig:rq1} shows the \textit{Tukey Test} with a small effect size, i.e., -0.02 to 0.01 in readability and -0.01 to 0.02 in understandability. The Top-2 code snippet shows overall better readability and understandability score than Top-1, Top-4 and Top-5 code snippets, as Top-3 shows better understandability score than Top-1 and Top-2 code snippets.

\vspace{-4mm}

\noindent
\begin{center}
\fbox{\begin{minipage}{25em}
\textbf{RQ \#1 Answer:} The ranking analysis confirms the \textbf{H \#1} hypothesis, i.e., a better-ranked code snippet is not necessarily more readable or understandable than a lower-ranked code snippet. The effect seems small, but we found Top-2 code snippets with overall better readability and Top-3 with overall better understandability score.
\end{minipage}}
\end{center}

\textbf{RQ \#2) How do general-purpose web search engines compare to each other in terms of readability and understandability features?}

Extracting the \textit{ANOVA} on the general-purpose web search engine independent variable, we obtained \textit{p-value = 1.207e-12} for readability and \textit{p-value = 0.0364} for understandability. This result means that there are significant differences between general-purpose web search engine code snippets in readability and understandability. The Figure \ref{fig:rq2} shows the \textit{Tukey Test} with  small effect, i.e., -0.02 to 0.02 in readability and -0.01 to 0.005 in understandability. In readability, the Google code snippets have overall better readability than Microsoft Bing and Yahoo!. In understandability, Google shows a small difference to Microsoft Bing.

\vspace{-4mm}

\noindent
\begin{center}
\fbox{\begin{minipage}{25em}
\textbf{RQ \#2 Answer:} Google has a better overall readability score compared to Microsoft Bing and Yahoo, but with a small effect. Google has a better overall understandability score compared to Microsoft Bing. 
\end{minipage}}
\end{center}

\textbf{RQ \#3) How do recommended sites containing code snippets compare to each other in terms of readability and understandability features?} 

Extracting the \textit{ANOVA} on recommended site independent variable, we obtained \textit{p-value = < 2.2e-16} for readability and \textit{p-value = < 2.2e-16} for understandability. This result means that there is significant differences between recommended sites code snippets in readability and understandability. The Figure \ref{fig:rq3} shows the \textit{Tukey Test} with a medium effect for readability, i.e., -0.15 to 0.10 and small effect on understandability, i.e.,  -0.04 to 0.08. In readability, \textit{geeksforgeeks} shows the higher overall score, and \textit{tutorialspoint} has overall better understandability.

\noindent
\begin{center}
\fbox{\begin{minipage}{25em}
\textbf{RQ \#3 Answer:} The recommended sites analysis confirms the \textbf{H \#3} hypothesis, i.e., there is significant differences between the readability and understandability score between the recommended sites. \textit{geeksforgeeks} has the overall best readability score, and \textit{tutorialspoint} has the overall best understandability score.
\end{minipage}}
\end{center}
\vspace{.4em}

\section{Discussion}



The \textit{geeksforgeeks} is a tutorial programming site containing code snippets with one comment per line of code, and high cohesion with one concept. These features contributes to produce a higher score on readability metric. For example, on the input query \textit{"How to append to a string?"}, the \textit{geeksforgeeks} code snippet \footnote{https://www.geeksforgeeks.org/java-program-to-add-characters-to-a-string/} had readability score = 0.94, and the \textit{tutorialspoint} code snippet \footnote{https://www.tutorialspoint.com/javaexamples/file\_append.htm/} has readability score = 0.44. The \textit{geeksforgeeks} code have more comments, and \textit{tutorialspoint} code snippet have more concepts in the same line, e.g. \textit{new BufferedWriter(new FileWriter("filename"))}, instead of \textit{geeksforgeeks} code snippet with one concept per line.

\begin{figure}[h]
\centerline{\includegraphics[width=0.27\textwidth]{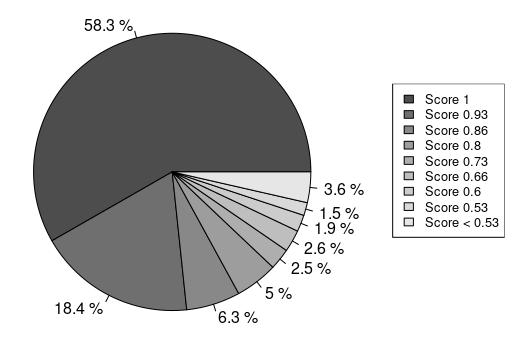}}
\vspace{-4mm}
\caption{Distribution of understandability score}
\label{fig:understandability_score}
\end{figure}

The understandability metric used in this research had a low effect in all analyses between independent variables. Figure \ref{fig:understandability_score} shows 58.3\% of the code snippets have maximum understandability scores. Many code snippets has few lines of \textit{API} calls, without \textit{if/else} conditions or \textit{for/while} loops. This result suggests that the understandability feature is more feasible to be used in complete classes from git repositories.

\vspace{-4mm}

\section{Threats to Validity}

\textbf{Number of sites to extract code snippets}: we selected five popular websites, which represent between 34\% and 38.1\% of the recommended sites for the input queries employed in this research. But the results could have variations if we increase the recommended sites.

\textbf{Multiple code snippets in same web page}: in this approach, we extracted the first code snippet of each recommended site (for StackOverflow, we extracted the code snippet from the accepted answer). The developer would test the first recommended code snippet on the web page. But some sites have more than one code snippet on the same web page, which a heuristic could be employed to extract readability and understandability scores from these multiple code snippets.  

\textbf{Readability and Understandability score precision}: we carefully selected the readability and understandability state-of-art tools, but their score could have some false positives/negatives, or even the readability metric score used in this research would not be reliable in the range [0,416, 0,600] \cite{Piantadosi2020}. 

\textbf{Queries modifications}: the addition of "example in java" tokens into each query could influence the web-search engines to produce their rankings, i.e., the query without modifications could produce a different ranking between the selected code snippets. An investigation about tokens additions is necessary for future works.

\textbf{Extract code snippets heuristics}: compared to the other sites, StackOverflow has a different heuristic on extracting code snippets, using the accepted answer. These heuristics require a sensitivity analysis for future works.

\section{Conclusions and Future Work}

In this empirical study, the recommended site independent variable has the highest effect on the readability score. Programming tutorial sites as \textit{geeksforgeeks} generally have code snippets in a specific format, but Q\&A sites as StackOverflow has many users sharing code snippets, which could lead to different code snippet formats. The readability standard deviation score on \textit{geeksforgeeks} is 0.11, and in StackOverflow is 0.20, which confirms more variance on Q\&A sites. The understandability feature have low effect in most of the scenarios, because most of them has few \textit{statements}, \textit{beacons} or \textit{motifs}. 

These results provide insights for future improvements. A qualitative study could be conducted to better understand the reasons for the variability in the readability and understandability score. A complementary study could include more sites, or even create a general regular expression to automatically extract code snippets from a large variety of sites. The general-purpose web search engines could be compared to specific code search engines, e.g., StackOverflow. Moreover, a heuristic could be proposed to evaluate the websites with multiple code snippets.

  \bibliographystyle{ACM-Reference-Format}
  \bibliography{acmart}

\end{document}